\documentstyle[11pt,newpasp,twoside,epsf]{article}
\newcommand{\Teff}{\hbox{$T_{\rm eff}$}}
\def\logg{$\log g$\thinspace}
\newcommand\etal{et al.\,}

\begin{document}
\title{Non-LTE effects in neutron star atmospheres}
 \author{K. Werner, J. Deetjen}
\affil{Institut f\"ur Astronomie und Astrophysik, Univ.\ T\"ubingen, Germany}

\begin{abstract}
We present the first non-LTE model atmospheres for neutron stars (NS). We study
their structure and NLTE effects on the emergent thermal radiation of old
isolated NS.
\end{abstract}


\noindent
Our models assume plane-parallel geometry, hydrostatic and radiative
equilibrium. The radiation transfer equations are solved simultaneously with
the NLTE rate equations by using an Accelerated Lambda Iteration technique
(Werner \& Dreizler 1999). Blanketing by millions of iron lines is considered
with an Opacity Sampling method. The models neglect magnetic fields, which is a
good approximation as long as B$< 10^8$--$10^{10}\, $G, depending on \Teff\ and
chemical composition (Zavlin \etal 1996). Iron opacities are calculated from
Kurucz line lists, and bound-free cross-sections from Opacity Project data.

We have computed a set of three models with different chemical composition:
pure H, He, and Fe models. We chose \Teff=200\,000\,K, \logg=14.39,
corresponding to a NS with M=1.4\,M$_\odot$ and R=10\,km. Fig.\,1 shows their
emergent fluxes. They display distinctly harder high energy tails than a
blackbody. Fig.\,2 shows details of the model structures. Due to NLTE effects,
temperatures show a slight increase in the surface layers. Departures from LTE
are most pronounced in the iron model. They affect the line formation
regions and hence fluxes in the strongest Fe line cores, typically of the order
10\%. The general spectral shape is unaffected. NLTE effects might be even
more important at higher \Teff.

\small
\vspace{2mm}
\noindent
Werner K., Dreizler S. 1999, Journal of Computational and Applied Math., in press\\
Zavlin V.E., Pavlov G.G., Shibanov Y.A. 1996, A\&A 315, 141
\normalsize

\begin{figure}[bth]
\epsfxsize=\textwidth
\epsffile{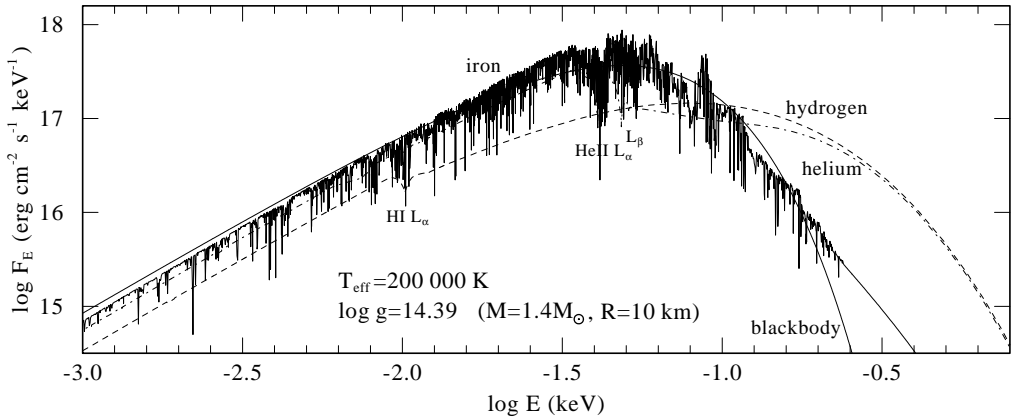}
\caption{
Emergent fluxes of NLTE neutron star model atmospheres
}
\end{figure}

\newpage

\begin{figure}[bth]
\epsfxsize=\textwidth
\epsffile{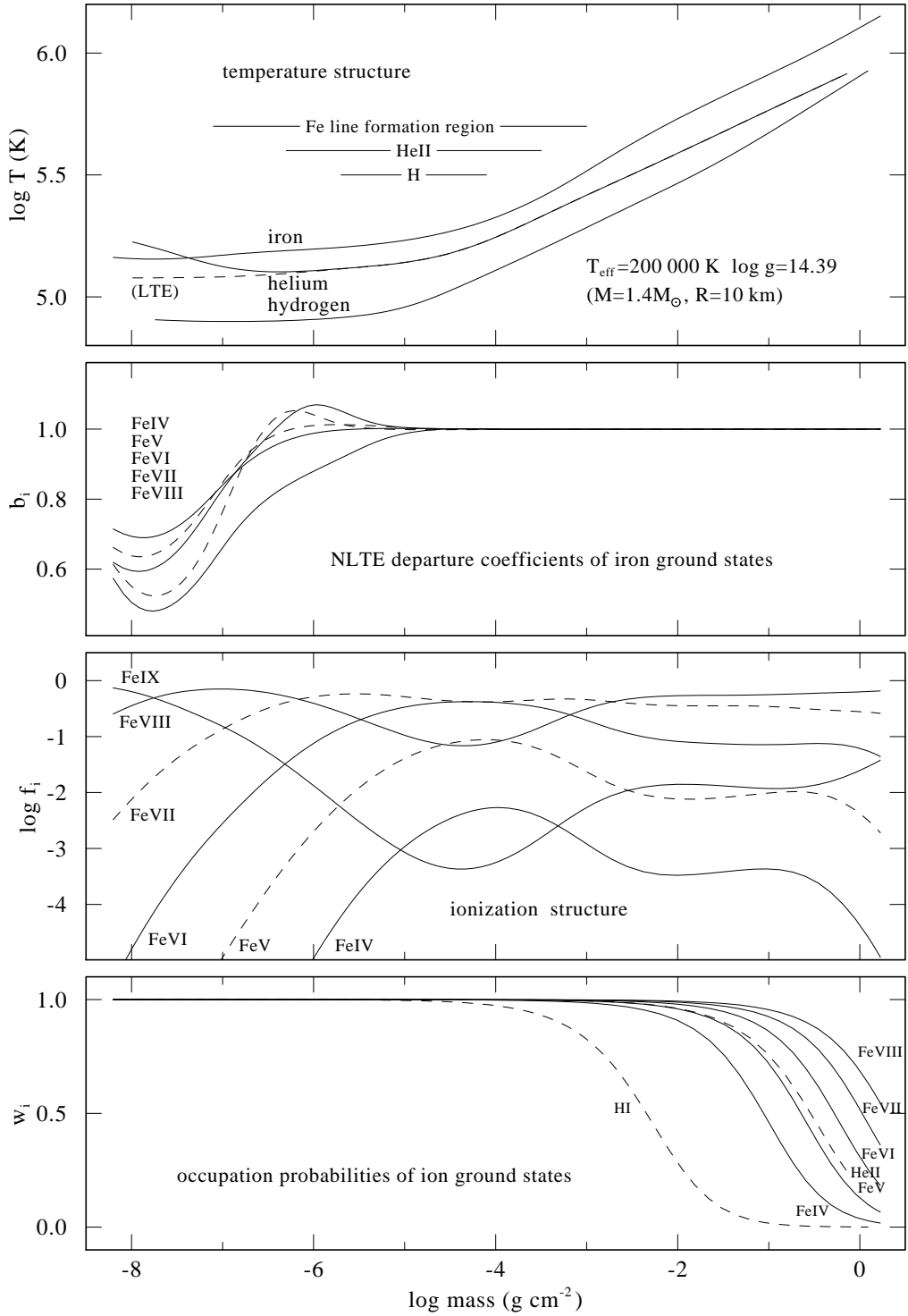}
\caption{
Structures of the NLTE neutron star model atmospheres. Departures from LTE
occur in the iron line formation region.
}
\end{figure}

\end{document}